# Masked Autoencoder Pretraining and BiXLSTM–ResNet Architecture for PET/CT Tumor Segmentation

Moona Mazher[1], Steven A Niederer[2] and Abdul Qayyum[2]

[1] Hawkes Institute, Department of Computer Science, University College London, London, United Kingdom
[2] National Heart and Lung Institute, Faculty of Medicine, Imperial College London, London, United Kingdom

**Abstract.** The accurate segmentation of lesions in whole-body PET/CT imaging is essential for tumor characterization, treatment planning, and response assessment, yet current manual workflows are labor-intensive and prone to inter-observer variability. Automated deep learning methods have shown promise but often remain limited by modality specificity, isolated time points, or insufficient integration of expert knowledge. To address these challenges, we present a two-stage lesion segmentation framework developed for the fourth AutoPET Challenge. In the first stage, a Masked Autoencoder (MAE) is employed for self-supervised pretraining on unlabeled PET/CT and longitudinal CT scans, enabling the extraction of robust modality-specific representations without manual annotations. In the second stage, the pretrained encoder is fine-tuned with a bidirectional XLSTM architecture augmented with ResNet blocks and a convolutional decoder. By jointly leveraging anatomical (CT) and functional (PET) information as complementary input channels, the model achieves improved temporal and spatial feature integration. Evaluation on the AutoPET Task 1 dataset demonstrates that self-supervised pretraining significantly enhances segmentation accuracy, achieving a Dice score of 0.582 compared to 0.543 without pretraining. These findings highlight the potential of combining self-supervised learning with multimodal fusion for robust and generalizable PET/CT lesion segmentation. Code will be available at https://github.com/RespectKnowledge/AutoPet_2025_BxLSTM_UNET_Segmentation

## 1    Introduction

The fourth AutoPET Challenge aims to explore interactive human-in-the-loop strategies for lesion segmentation in two distinct tasks: (1) whole-body PET/CT lesion analysis and (2) longitudinal CT lesion tracking [1].

Hybrid imaging modalities, particularly Positron Emission Tomography combined with Computed Tomography (PET/CT), are central to the clinical management of patients with malignant tumors. Currently, treatment response is assessed by radiologists through the manual identification and measurement of lesions across serial scans. This process is labor-intensive, subject to inter-observer variability, and reduces the wealth of information embedded in volumetric imaging to simple one-dimensional measurements. As a result, valuable morphologic and metabolic information is often overlooked, despite its potential prognostic relevance [2].

Automated lesion detection and segmentation methods offer a path toward faster, more comprehensive, and reproducible tumor characterization. Deep learning-based models, in particular, have shown promise, but many existing approaches are restricted to single modalities, isolated time points, or highly specific imaging settings, limiting their generalizability. Furthermore, the integration of human expertise into the model development or inference process has not been sufficiently explored, despite its potential to enhance robustness and clinical usability. There are a lot of segmentation models proposed in medical imaging [3-12].

To advance the state of the art, AutoPET/CT IV provides multimodal datasets, including PET/CT and longitudinal CT series, along with varying levels of annotation. The challenge encourages participants to explore novel architectures, foundation model integration, and data-centric strategies to improve lesion segmentation and tracking.
In this work, we propose a bidirectional XLSTM-based architecture with a residual encoder and a conventional convolutional decoder for lesion segmentation. Our method is designed as a two-stage pipeline. In the first stage, we employ a Masked Autoencoder (MAE) [3, 13] for self-supervised representation learning using the unlabeled challenge data, which enables the network to capture modality-specific features from both PET and CT scans without requiring manual annotations. In the second stage, we fine-tune the pretrained encoder using a bidirectional XLSTM augmented with a ResNet backbone to improve temporal and spatial feature integration for Task 1. Furthermore, we fuse CT and PET modalities as complementary input channels, allowing the model to simultaneously exploit anatomical and functional information for robust lesion detection and segmentation.

## 2 Proposed Method

### 2.1 Task 1: Whole-body PET/CT

For Task 1, we used the FDG and PSMA PET/CT cohorts. The FDG dataset includes 1,014 studies from 900 patients (melanoma, lymphoma, or lung cancer) plus 513 negative controls. The PSMA dataset comprises 597 studies from 378 prostate cancer patients. Scans were acquired at multiple sites using different PET/CT scanners, with PET and CT anatomically aligned [1].

Each case contains a PET volume, a corresponding CT volume, and a binary lesion mask manually annotated by expert radiologists. Annotations followed a standardized protocol: (1) tracer-avid lesion identification using PET and CT information, and (2) manual segmentation on axial slices. This dataset was used for our two-stage training: Stage 1 pretraining with a masked autoencoder on unlabeled data, followed by Stage 2 fine-tuning with our BiXLSTM–ResNet encoder and convolutional decoder using the provided annotations.

## 2.2 Task 2: Longitudinal CT Screening

For Task 2, we accessed 300 melanoma patients with baseline and follow-up whole-body CT scans. Each case included CT volumes and integer-coded lesion masks annotated by two radiologists. Although we did not participate in Task 2, these scans were incorporated only for self-supervised pretraining to improve model generalization.

## 2.3 Data and Preprocessing

We trained our proposed model using the autoPET Task 1 dataset, which includes 1,014 FDG-PET/CT studies and 597 PSMA-PET/CT studies. Preprocessing consisted of:

- a. Resampling PET and CT volumes to the same voxel resolution.
- b. Intensity normalization of PET (SUV scale) and CT volumes.
- c. nnUNet-compatible NIfTI formatting, including training/validation splits from *splits_final.json*.

For data augmentation during training, we applied standard nnUNet strategies:

- a. **Spatial augmentations:** random rotation, flipping, scaling, elastic deformation.
- b. **Intensity augmentations:** random brightness and gamma adjustments.

Table 1. Algorithm details

| Team name | Dolphins |
|---|---|
| **Algorithm name** | BiXLSTM–ResNet UNet |
| **Data pre-processing** | Normalization, Resampling to common voxel size, nnUNet-compatible NIfTI |
| **Data post-processing** | None |
| **Training data augmentation** | Random rotation, flipping, scaling, elastic deformation, brightness, gamma |
| **Test-time augmentation / Ensembling** | Flipping, mirroring; 5-fold cross-validation |
| **Standardized framework** | nnUNet v2 |
| **Network architecture** | BiXLSTM encoder + ResNet blocks, convolutional decoder |
| **Loss** | Dice + Cross-Entropy |
| **Training data (size)** | 1014 FDG + 597 PSMA PET/CT |
| **Dimensionality / Patch size** | 128×192×160 |
| **Use of pre-trained models** | Masked Autoencoder pretraining on unlabeled Task 1 & Task 2 data |
| **GPU hardware** | 1 × NVIDIA A6000 |

Test-time augmentation consisted of flipping and mirroring, and we used 5-fold cross-validation to enhance robustness. All training and validation were conducted within the nnUNet v2 framework [14], adapted for our BiXLSTM–ResNet encoder and convolutional decoder. Training was performed on a single NVIDIA A6000 GPU, with patch size 3D: 128×192×160. No external pretrained models were used aside from our masked autoencoder pretraining on unlabeled PET/CT and longitudinal CT scans.

Table 1 provides an overview of our BiXLSTM–ResNet UNet model and training pipeline. The model was pretrained with a self-supervised MAE on unlabeled PET/CT and longitudinal CT data, then fine-tuned on annotated Task 1 PET/CT scans with standard nnUNet preprocessing, augmentation, and 5-fold cross-validation.

## 2.4 Algorithm / Model

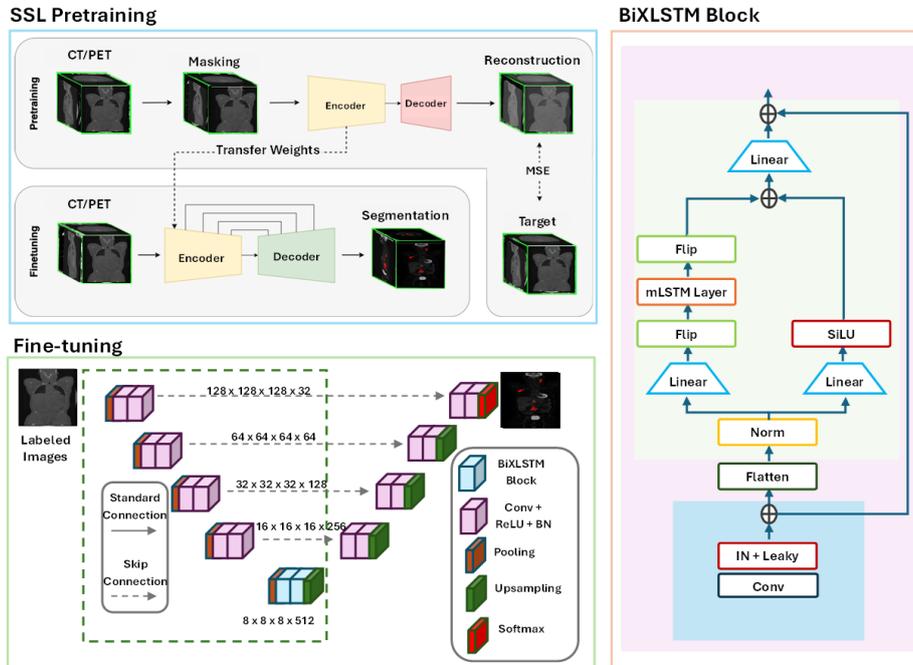

**Figure 1.** Overview of the two-stage PET/CT lesion segmentation pipeline. Stage 1 employs a Masked Autoencoder (MAE) for self-supervised pretraining on unlabeled PET/CT and longitudinal CT scans to learn robust modality-specific features. Stage 2 fine-tunes the pretrained encoder with a BiXLSTM–ResNet architecture and a convolutional decoder on annotated PET/CT data, fusing PET and CT channels for voxel-wise lesion segmentation.

Our method follows a two-stage training pipeline:

**Stage 1 –** Self-supervised Pretraining: We employ a Masked Autoencoder (MAE) [13] on unlabeled PET/CT scans from Task 1 and longitudinal CT scans from Task 2. This allows the network to learn robust feature representations from both modalities without requiring manual annotations. The MAE reconstructs masked patches of the input volume, encouraging the model to capture global and local anatomical and metabolic structures.

**Stage 2** – Fine-tuning for Lesion Segmentation: The pretrained encoder is integrated with a bidirectional XLSTM [15] enhanced with ResNet blocks, paired with a standard convolutional decoder. The network is fine-tuned using the annotated Task 1 PET/CT dataset to perform voxel-wise lesion segmentation. Both CT and PET channels are fused as input, enabling the model to jointly leverage anatomical and functional information. Training optimizes a combination of Dice loss and cross-entropy loss, with standard nnUNet data augmentation and 5-fold cross-validation to improve generalization.

This two-stage approach allows the model to first learn modality-specific representations and then adapt these features for accurate lesion segmentation, achieving improved performance compared to training from scratch.

## 3      Results

Our proposed model has been used for training, and our score on the leaderboard is explained in Table 2.

Table 2. Performance of Proposed Models on autoPET Task 1

| Model Description | Dice Score (mean) | False Negative Volume (mean, cm³) | False Positive Volume (mean, cm³) |
|---|---|---|---|
| BiXLSTM–ResNet (no SSL, trained from scratch) | 0.543 | 23.26 | 15.05 |
| BiXLSTM–ResNet_Enc + SSL | 0.580 | 13.78 | 15.43 |
| BiXLSTM–ResNet _Bot + SSL | 0.582 | 15.08 | 13.69 |

We evaluated three variants of our BiXLSTM–ResNet model on the AutoPET Task 1 PET/CT dataset to assess the effect of self-supervised pretraining on segmentation performance. The baseline model, trained from scratch without self-supervised learning, achieved a mean Dice score of 0.543. This score indicates moderate agreement with the ground truth lesion masks, but the relatively high false negative volume of 23.26 cm³ shows that a considerable number of lesions were not detected. The false positive volume of 15.05 cm³ indicates that some non-lesion regions were incorrectly classified as lesions.

When self-supervised MAE pretraining was added, followed by supervised fine-tuning on annotated Task 1 data, the model demonstrated improved performance. The Dice score increased to 0.582, reflecting better overlap with the ground truth masks. The false negative volume decreased to 15.08 cm³, indicating that more lesions were correctly identified, and the false positive volume was reduced to 13.69 cm³, showing that predictions became more precise and consistent.

Overall, this comparison illustrates that self-supervised pretraining significantly enhances the model's ability to capture relevant features from PET and CT scans, while task-specific fine-tuning further refines segmentation accuracy. These improvements demonstrate the importance of combining unsupervised representation learning with annotated data to achieve robust and reliable lesion segmentation.

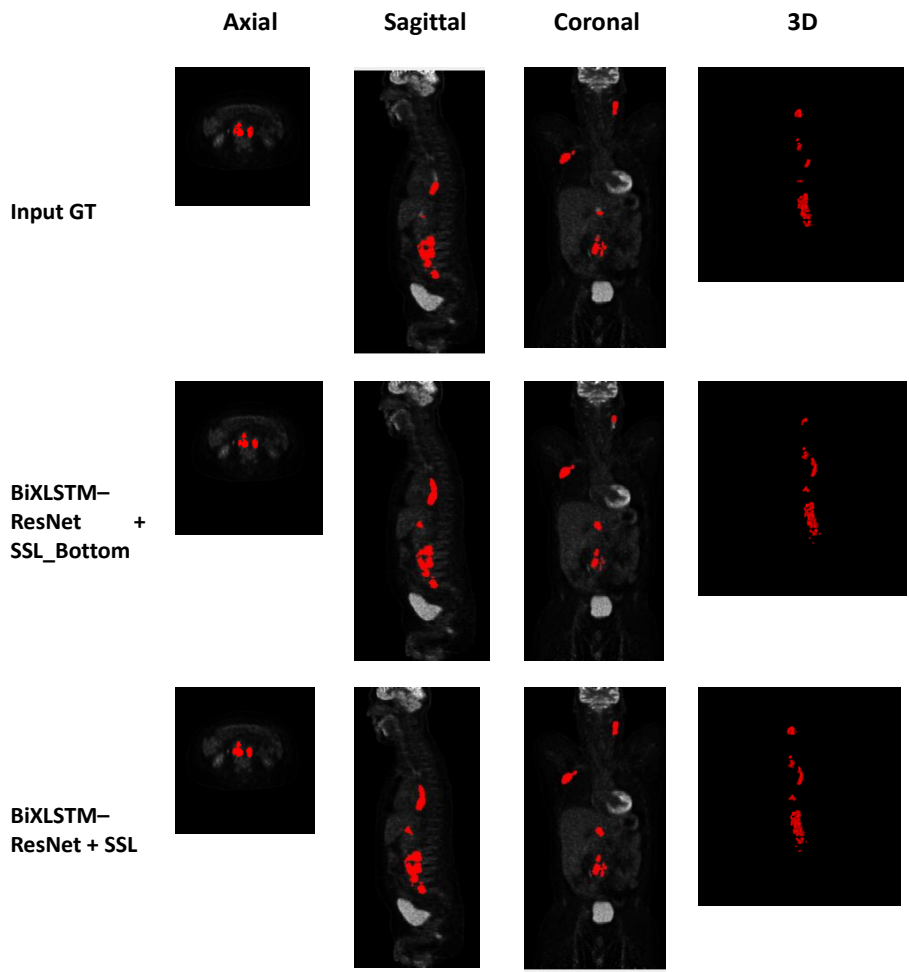

**Figure 2.** Visualization of whole-body tumor segmentation results in PET. Examples are shown in axial, sagittal, coronal, and 3D views, comparing the ground truth (GT) with predictions from the proposed BiXLSTM–ResNet model.

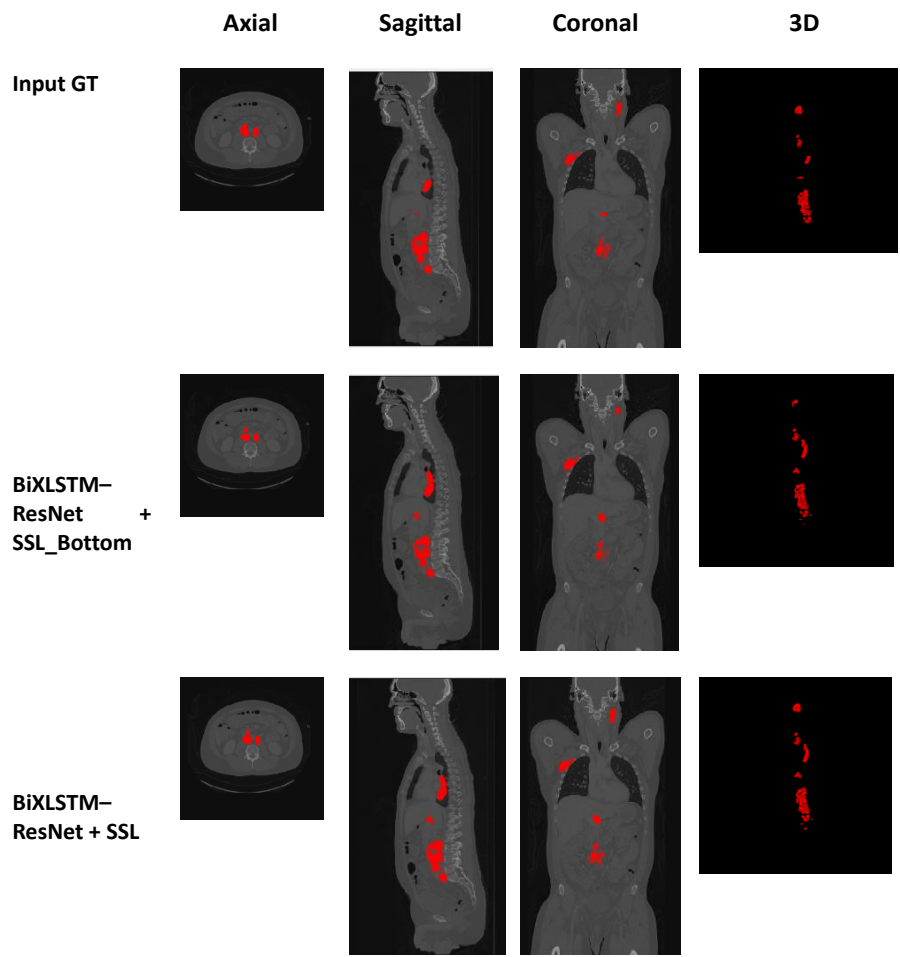

**Figure 3.** Visualization of whole-body tumor segmentation results in CT. Examples are shown in axial, sagittal, coronal, and 3D views, comparing the ground truth (GT) with predictions from the proposed BiXLSTM–ResNet model.

Figures 2 and 3 illustrate the segmentation performance of our proposed BiXLSTM–ResNet model with self-supervised MAE pretraining on PET and CT volumes, respectively. In the PET visualization (Figure 2), the ground truth lesion masks and model predictions are shown across axial, sagittal, coronal, and 3D views. The results demonstrate that the model accurately identifies tracer-avid lesions, effectively reducing false negatives and closely aligning with expert annotations. Similarly, the CT visualizations (Figure 3) highlight the model's ability to delineate anatomical boundaries when fused with PET information, capturing lesions even in regions with subtle structural contrast. Overall, these figures confirm that self-supervised pretraining improves both spatial and volumetric accuracy, consistent with the quantitative performance metrics, including a mean Dice score of 0.582, reduced false negative volume (15.08 cm³), and lower false positive volume (13.69 cm³).

## 4 Discussion

In this study, we presented a two-stage framework for PET/CT lesion segmentation that leverages self-supervised Masked Autoencoder (MAE) pretraining and a BiXLSTM–ResNet architecture. The quantitative results on the AutoPET Task 1 dataset demonstrate that self-supervised pretraining substantially improves segmentation performance. Specifically, the baseline model trained from scratch achieved a Dice score of 0.543, with high false negative and false positive volumes, while the model incorporating MAE pretraining achieved a higher Dice score of 0.582, along with reduced false negative (15.08 cm³) and false positive (13.69 cm³) volumes. These results highlight the ability of self-supervised learning to extract robust modality-specific features from unlabeled PET/CT and longitudinal CT data, which significantly enhances downstream lesion segmentation.

The fusion of PET and CT inputs allows the network to jointly leverage functional and anatomical information, resulting in more precise lesion delineation, especially for small or complex lesions that may be difficult to detect with a single modality. The bidirectional XLSTM further strengthens the model by capturing spatial and temporal context across slices, which is crucial in whole-body imaging where lesions are distributed across multiple regions. Qualitative visualizations confirm that the model accurately identifies lesion boundaries and reduces both under- and over-segmentation errors.

These findings emphasize the broader potential of self-supervised learning in medical imaging, particularly in domains where labeled data are limited or costly to obtain. By pretraining on large unlabeled datasets, models can learn generalizable feature representations that improve accuracy and robustness. Future work could extend this approach to longitudinal lesion tracking, multi-tracer PET studies, or integration with clinical data to provide comprehensive tumor characterization. Overall, our framework demonstrates a promising path toward automated, reliable, and clinically applicable PET/CT lesion segmentation.

## 5 Conclusion

In this work, we proposed a two-stage deep learning framework for PET/CT tumor segmentation, combining self-supervised pretraining with a Masked Autoencoder and a BiXLSTM–ResNet architecture. By first learning robust modality-specific representations from unlabeled PET/CT and longitudinal CT scans, and then fine-tuning on annotated data, our approach effectively integrates anatomical and functional information while capturing temporal and spatial dependencies. Evaluation on the AutoPET Task 1 dataset demonstrated that self-supervised pretraining significantly improves segmentation performance, achieving a mean Dice score of 0.582 and reducing false negative and false positive volumes compared to training from scratch. These results highlight the potential of combining self-supervised learning with multimodal fusion for accurate, generalizable, and clinically relevant lesion segmentation, paving the way for more efficient and reliable tumor characterization in whole-body PET/CT imaging.